\documentclass[12pt]{iopart}
\usepackage{iopams}

\begin{document}

\title{Bounce and wormholes}

\author{N Pinto-Neto, F P Poulis and J M Salim}
\address{Centro Brasileiro de Pesquisas F\'{i}sicas, Rua
Xavier Sigaud, 150, CEP 22290-180, Rio de Janeiro, Brazil.}
\eads{\mailto{nelson.pinto@pq.cnpq.br}, \mailto{fppoulis@cbpf.br}, \mailto{jsalim@cbpf.br}}

\begin{abstract}
We investigate if theories yielding bouncing cosmological models
also generate wormhole solutions. We show that two of them present
sensible traversable static wormhole solutions, while for the third possibility
such solutions are absent.
\end{abstract}


\maketitle

\section{Introduction}
With the observation of the present acceleration of the Universe \cite{acc}, it seems that fluids that violate the strong energy condition (SEC) \cite{null} must indeed exist in Nature, and anti-gravitate. It is also speculated that such acceleration might also be driven by yet more exotic fluids, called phantom fields, which violate the weak energy condition (WEC), and may cause a future Big-Rip singularity \cite{rip}. These fluids can also
be effectively obtained through non-minimal couplings between gravity and other fields \cite{nonm},
quantum corrections \cite{qc}, or non-standard interactions among ordinary fields satisfying all the energy
conditions \cite{be}.

Such kind of fluids may also play a fundamental role in the early Universe, either yielding an
inflationary period \cite{inflation}, and/or avoiding the initial singularity as in the bouncing
models \cite{bounce}. Regarding the later possibility, it is also known that the violation of SEC
may not be sufficient to produce a bounce in the past: some period of WEC violation is also required if
the spatial sections have not positive curvature \cite{PP}.

Fluids violating WEC are also necessary to obtain wormhole solutions \cite{thorne} hence,
in view of all those scenarios in which exotic sources are seriously taken into account, we can
investigate in which of them traversable wormhole solutions can be obtained.

In the present paper we investigate the existence of wormhole
solutions in theoretical frameworks which produce bouncing
cosmological models. The existence and properties of these
cosmological bounces have been studied in many
papers \cite{Novello,Novello1,PP2,Novello2}, but it remains to be
investigated if such theoretical frameworks can produce wormhole
solutions, which is a natural expectation to have due to the
enormous parallel between bounces and wormholes (which are nothing
but bounces in space).

Our work will be based in three models of bouncing universes. The
first one considers a non-minimal coupling between electromagnetic
and gravitational fields caused by quantum electrodynamics
corrections to general relativity \cite{Novello,Drummond}, which could be
related to magnetic fields observed in galaxies and intergalactic
medium \cite{prokopec,turner}. This coupling leads to non-linearities in
the theory and, consequently, violations of the energy conditions
that enabled not only a bounce \cite{Novello} but also a very nice
traversable wormhole solution.

The next one considers a model of universe in a Weyl Integrable
Space-time (WIST) \cite{Novello1}, yielding an effective negative
energy scalar field, which evolves from a Minkowski
configuration in the infinite past, goes through a bounce and
finish in another asymptotic flat universe. This pure geometric
model can also give rise to a perfectly reasonable traversable wormhole.

The last case treated involves non-linear corrections to the electromagnetic field. Such corrections are needed when very intense fields are considered, where the particle creation phenomenon occurs. These non-linearities leads to the violation of energy conditions and allow a bounce \cite{Novello2} but not a wormhole. We will show that whenever one imposes the boundary conditions that characterizes the wormhole throat, the asymptotic behavior of the spacetime solution becomes unacceptable.
Therefore, we have an example of a theoretically acceptable source that violates WEC and do not lead to a wormhole, showing that this condition is necessary but not sufficient to obtain such type of geometries.

Finally, we conclude that, although we have explicitly shown examples of wormholes solutions obtained using sources already present in other frameworks, this is not enough to consider such type of configuration as real
as long as they are static. This still depends on dynamical treatments, which are beyond the scope of this paper.

\section{Non-minimal coupling between electromagnetic and gravitational
fields}

Non-minimal coupling between electromagnetic and
gravitational fields appears when one takes into account the vacuum
polarization influence on the propagation of photons in a gravitational
background using the one-loop approximation \cite{Drummond}. In the present section, we will consider a non-minimal coupling term added to Einstein-Maxwell action for gravity. It was used to obtain a non singular cosmological model \cite{Novello}, to study the production of primordial magnetic fields \cite{prokopec,turner}
and to yield inflation \cite{slava}. The complete lagrangian reads

\begin{equation}
L=\sqrt{-g}\left[\frac{1}{2k}(1+ \lambda A _{\mu}A ^{\mu}) R
-\frac{1}{4}F_{\mu\nu}F^{\mu\nu}\right],\label{nmc1}
\end{equation}

\noindent where $k=8\pi G$, $G$ is the Newton's constant, $\lambda$ is the
non-minimal coupling constant, and $F^{\mu\nu}$ is the usual Maxwell
tensor constructed from $A^{\mu}$. The
variation of this Lagrangian with respect to $g_{\mu\nu}$ and $A^{\mu}$
yields the following equations:

\begin{equation}
\left(1+\lambda A^{2}\right)G_{\mu\nu}+\lambda\square
A^{2}g_{\mu\nu}-\lambda\nabla_{\nu}\nabla  _{\mu}A^{2}+\lambda
RA_{\mu}A_{\nu}=kT_{\mu\nu}, \label{eq de mov var gmn ANM}
\end{equation}

\noindent and

\begin{equation}
F^{\mu\nu}\,_{||\nu}=-\frac{\lambda}{k}RA^{\mu}, \label{eqmax}
\end{equation}

\noindent where $G_{\mu\nu}$ is the usual Einstein tensor, $A^2\equiv A^{\mu}A_{\mu}$, and
$T_{\mu\nu}$ is the Maxwell stress-energy tensor,

\begin{equation}
T_{\mu\nu}=
F_{\mu\alpha}F^{\alpha}_{\nu}+\frac{1}{4}g_{\mu\nu}F_{\alpha\beta}F^{\alpha\beta}.
\end{equation}

We will look for solutions with the same source used in the
cosmological solution obtained in Ref. \cite{Novello}, which presents a bounce,
where $A_{\mu}= w_{|\mu}$. With this choice for the vector potential, it follows
from \eref{eqmax} that $R=0$, and  the set of equations reduces to

\begin{equation}
R_{\mu\nu} = \frac{\Upsilon _{|\mu||\nu}} {\Upsilon },\label{eqc1}
\end{equation}

\begin{equation}
g^{\mu\nu}\Upsilon _{|\mu||\nu}=0, \label{eqc2}
\end{equation}

\noindent in which the last one comes from the trace of \eref{eq de mov var gmn ANM} and we have introduced the new variable $\Upsilon \equiv 1+\lambda A^{2}$.

We are interested in a static spherical geometry, with metric given by

\begin{equation}
ds^{2}=-e^{2\phi(r)}dt^{2}+\frac{dr^{2}}{\left(1-\frac{b(r)}{r}\right)}+r^{2}d\Omega^{2}.
\label{metrica Escalar}
\end{equation}

The field equations \eref{eqc1} and \eref{eqc2}, in this case, reduce to

\begin{eqnarray}
-\left(1-\frac{b}{r}\right)\frac{\phi'\Upsilon '}{\Upsilon }  =  \frac{b'}{r},\label{Sis ANM ro}\\
\frac{1}{\Upsilon }\sqrt{1-\frac{b}{r}}\left(\sqrt{1-\frac{b}{r}}\Upsilon ' \right)'  =  -\frac{b}{r^3}+2\left(1-\frac{b}{r}\right)\frac{\phi'}{r},\label{Sis ANM p}\\
\left(1-\frac{b}{r}\right)\frac{\Upsilon '}{r\Upsilon }  = 
\left(1-\frac{b}{r}\right)\left[\phi''+\phi'\left(\phi'+\frac{1}{r}\right)\right]-\frac{\left(b'r-b\right)}{2r^{2}}\left(\phi'+\frac{1}{r}\right),\label{Sis ANM tau}\\
\frac{\sqrt{1-\frac{b}{r}}}{r^{2}e^{\phi}}\left(r^{2}e^{\phi}\sqrt{1-\frac{b}{r}}\Upsilon '\right)' = 0.\label{Dalam
Omega}
\end{eqnarray}

\noindent This last one implies that

\begin{equation}
r^{2}e^{\phi}\sqrt{1-\frac{b}{r}}\Upsilon '=D
\label{constD},
\end{equation}

\noindent a constant, so that \eref{Sis ANM p} can be rewritten as

\begin{equation}
\frac{D}{\Upsilon }\sqrt{1-\frac{b}{r}}\left(\frac{1}{r^{2}e^{\phi}}\right)' = -\frac{b}{r^3}+2\left(1-\frac{b}{r}\right)\frac{\phi'}{r}.
\end{equation}

A simple exact solution to this non-linear system of
differential equations reads

\begin{eqnarray}
b(r)= r_{0},\label{Sol b ANM}\\
\phi(r)= \phi_{0},\label{Sol phi ANM}\\
\Upsilon (r)=
\frac{2D}{r_{0}e^{\phi_{0}}}\sqrt{1-\frac{r_{0}}{r}}.\label{Sol Omega ANM}
\end{eqnarray}

Let us now examine the properties and traversability of this wormhole,
following the lines of \cite{thorne}. The embedding function $z(r)$ reads

\begin{equation}
z(r)=2r_0\left(\frac{r}{r_0}-1\right)^{1/2}
\label{emb1}
\end{equation}

\noindent showing that it is a parabolic wormhole. The proper distance is given by

\begin{equation}
l(r)=\sqrt{r}\sqrt{r-r_0}-\frac{r_0}{2}\ln\left(\frac{\sqrt{r}-\sqrt{r-r_0}}{\sqrt{r}+\sqrt{r-r_0}}\right),
\label{prop}
\end{equation}

\noindent which tends towards $r$ when $r$ is big.

As long as $\phi$ is constant, one can traverse the wormhole with constant speed,
and the tidal radial acceleration is null. For the tidal lateral acceleration to be less than
one Earth gravity, one has the condition on the speed $v$

\begin{equation}
\frac{r_0\gamma^2}{2r^3}\frac{v^2}{c^2} \geqslant \frac{1}{(10^{10}{\rm cm})^2},
\label{conv}
\end{equation}

\noindent where $\gamma$ is the usual relativistic $\gamma$-factor, which we are assuming to be approximately equal to one, since we are considering non-relativistic velocities. Its maximum value is at the minimum of the throat at $r_0$, which implies that

\begin{equation}
v \leqslant 42 {\rm m/s} \frac{r_0}{10 {\rm m}}.
\label{convb}
\end{equation}

If the trip through the wormhole begins and ends in stations where the curvature
of the wormhole is negligible, with deviations, say, of $0.01\%$ from flatness,
then they must be located at coordinate distances of order $r\approx 10^4 r_0$,
yielding a total proper distance of $\Delta l \approx 2\times 10^4 r_0$. This
gives a total time travel of

\begin{equation}
\Delta t = \frac{\Delta l}{v} \geqslant 1 {\rm h} 23 {\rm min}.
\label{cont}
\end{equation}

\noindent For a one year travel one needs

\begin{equation}
v \approx 6.25 r_0 10^{-4}/{\rm s}.
\label{convy}
\end{equation}

Note that the electromagnetic energy required to construct this
wormhole is null because $F_{\mu\nu}$, and hence $T_{\mu\nu}$, are
null.

\section{Effective negative energy scalar field}

As mentioned in the Introduction, cosmological bounces can also be
induced by negative energy fluids \cite{FPP,PP2,BV}. We will consider
in this section a negative energy stiff matter fluid represented
by a massless free scalar field, which can appear due to interactions
among positive energy fluids \cite{be}, or through a pure geometrical model
where space-time is represented by a manifold with a geometric structure determined by
the Weyl geometry  \cite{Novello1}. In this later case, the
model is empty, it starts as a flat Minkowski space-time which, through
instabilities, begin to collapse into a homogeneous and isotropic
Weyl geometry.

The dynamics of the model is obtained from the action

\begin{equation}
S=\int\sqrt{-g}\left(\hat{R}+\xi\hat{\nabla}_{\alpha}W^{\alpha}\right),
\end{equation}

\noindent where $\hat{R}$ is the Ricci tensor in the Weyl geometry, $W^{\alpha}$ is the Weyl vector
satisfying $\hat{\nabla}_{\alpha}g_{\mu\nu}=W_{\alpha}g_{\mu\nu}$,
and
$\hat{\nabla}_{\alpha}$ is the covariant derivative with the Weyl
affinity.

Taking a Weyl integrable space-time, where the Weyl vector is a gradient given by
$W_{\lambda}=\varphi _{|\lambda}$, and performing the variation of the action $S$ with respect to the pair
$(g_{\mu\nu}, \varphi) $ of independent variables yields the following equations of motion in terms
of Riemannian geometrical quantities (see \cite{Novello1} for details):

\begin{equation}
G_{\mu\nu}= - \lambda^{2}(\varphi _{|\mu}\varphi _{|\nu}-\frac{1}{2}\varphi _{|\alpha}\varphi ^{|\alpha}g_{\mu\nu})
\equiv - \lambda^{2}T^{\varphi}_{\mu\nu},
\label{Eq Einstein WIST}
\end{equation}

\begin{equation}
\square\varphi =0, \label{KG WIST}
\end{equation}

\noindent in which $\lambda^{2}=(4\xi-3)/2$. Note that if this constant is positive, one has an effective negative energy fluid, which gives rise to cosmological bounces \cite{Novello1,PP2,FPP,BV}.

We are looking for the
possibility of a static wormhole solution for the above set of
equations. Using the spherical geometry given by the metric \eref{metrica Escalar}, the equations \eref{Eq Einstein WIST} and \eref{KG WIST}, after a simple manipulation, read

\begin{eqnarray}
b' = -\frac{\lambda^{2}}{2}\left(1-\frac{b}{r}\right)\left(\varphi '\right)^{2}r^{2},\label{bl escalar}\\
\phi' = \frac{b-\frac{r^{3}\lambda^{2}}{2}\left(1-\frac{b}{r}\right)\left(\varphi '\right)^{2}}{2r\left(r-b\right)},\label{phil Escalar}\\
\left[\frac{1}{2}\left(1-\frac{b}{r}\right)\left(\varphi '\right)^{2}\right]'
 =  -\left(1-\frac{b}{r}\right)\left(\varphi
'\right)^{2}\phi'-\frac{2}{r}\left(1-\frac{b}{r}\right)\left(\varphi '\right)^{2}.\label{pl
Escalar}
\end{eqnarray}

\noindent This last one can be rewritten as

\begin{equation}
\left[\left(r^{2}e^{\phi}\right)^{2}\left(1-\frac{b}{r}\right)\left(\varphi '\right)^{2}\right]'=0.
\end{equation}

Now it is easy to integrate the resulting system of equations. The
solution is:

\begin{equation}
\varphi(r)=-\frac{\sqrt{2}}{\lambda}\arctan\left(\frac{r_{0}}{\sqrt{r^{2}-r_{0}^{2}}}\right)+ \frac{\pi\sqrt{2}}{2\lambda} +\varphi
(r_{0}), \label{fi(r)Escalar}
\end{equation}

\begin{equation}
 b(r)=\frac{r^2_0}{r},
\end{equation}

\begin{equation}
 \phi(r)=\phi_0.
\end{equation}

The space-time generated by this solution is represented by the geometry described by the following metric:

\begin{equation}
ds^{2}=-e^{2\phi_{0}}dt^{2}+\frac{dr^{2}}{\left(1-\frac{r_{0}^{2}}{r^{2}}\right)}+r^{2}d\Omega^{2}.
\label{metrica Escalar2}
\end{equation}

As it was done in the previous section, let us examine the properties and traversability
of this wormhole, again following the lines of \cite{thorne}. The embedding function $z(r)$ now reads

\begin{equation}
z(r)=r_0{\rm arccosh}\left(\frac{r}{r_0}\right),
\label{emb2}
\end{equation}

\noindent showing that it is a hyperbolic wormhole. The proper distance is given by

\begin{equation}
l(r)=r\sqrt{1-\left(\frac{r_0}{r}\right)^2},
\label{prop2}
\end{equation}

\noindent which, again, tends towards $r$ when $r$ is big. Also, as long as $\phi$ is constant, one can traverse the wormhole with constant speed, and the tidal radial acceleration is null. For the tidal lateral acceleration to be less than one Earth gravity, one has the condition on the maximum speed $v$ at $r=r_0$ inside the throat,

\begin{equation}
v \leqslant 30 {\rm m/s} \frac{r_0}{10 {\rm m}}.
\label{convb2}
\end{equation}

For a $0.01\%$ departure from flatness at the stations,
they must be located at coordinate distances of order $r\approx 10^2 r_0$,
yielding a total proper distance of $\Delta l \approx 2\times 10^2 r_0$. This
gives a total time travel of

\begin{equation}
\Delta t = \frac{\Delta l}{v} \geqslant 67 {\rm s}.
\label{cont2}
\end{equation}

\noindent For a one year travel one needs

\begin{equation}
v \approx 6.25 r_0 10^{-6}/{\rm s}.
\label{convy2}
\end{equation}

In this case, some (negative) energy for the scalar field is
required to construct this wormhole as long as its energy momentum
tensor $T^{\varphi}_{\mu\nu}$ is not null. The energy density is
$\rho = -r_0^2/r^4$, which in the throat at $r=r_0$ reaches its
maximum value $\rho = -1/r_0^2$, and the total (negative) mass of
the scalar field reads $M=-2\pi^2 r_0$. Hence, smaller values of
$r_0$ requires less mass but bigger densities inside the throat.

\section{Non-linear electrodynamics}

In \cite{Novello2} bouncing solutions where obtained considering a generalization of Maxwell electrodynamics, given by non-linear local covariant and gauge-invariant terms which depend on field invariants up to second order, as the source of classical Einstein's equations. This modification is expected to be relevant when the fields reach large values, as occurs in the early universe. One expects that such modifications may also be important inside the hypothetical throat of a wormhole, generated by this source.

The generalization of Maxwell electrodynamics, up to second order, is determined by the following lagrangian \cite{Novello2}

\begin{equation}
L=-\frac{1}{4}F+\alpha F^{2}+\beta G^{2},
\label{LagrangeanaEMnlinear}
\end{equation}

\noindent where

\[
F :=F_{\mu\nu}F^{\mu\nu},
\]

\noindent and

\[
G:=\frac{1}{2}\eta_{\alpha\beta\mu\nu}F^{\alpha\beta}F^{\mu\nu},
\]

\noindent where $\alpha$ and $\beta$ are arbitrary constants and $\eta_{\alpha\beta\mu\nu}$ is a totally anti-symmetric tensor. According to that same reference, in the early universe, the source should be identified with a hot primordial plasma. And, to keep isotropy, spatial average should be taken into account \cite{Tolman}. This can be done by associating the following mean values to the electric ({\boldmath $E$}) and magnetic ({\boldmath $H$}) fields.

\begin{eqnarray}
\overline{E_{i}}=0,\qquad \overline{H_{i}}=0,\qquad \overline{E_{i}H_{j}}=0,\label{medias E H EH}\\
\overline{E_{i}E_{j}}=-\frac{1}{3}E^{2}g_{ij},\label{media EE}\\
\overline{H_{i}H_{j}}=-\frac{1}{3}H^{2}g_{ij},\label{media HH}
\end{eqnarray}

\noindent according to the definition

\begin{equation}
\overline{C}\equiv\lim_{V\rightarrow V_{0}}\frac{1}{V}\int C\sqrt{-g}d^{3}x^{i};\label{def media}
\end{equation}

\noindent to an arbitrary quantity $C$, in which $V=\int\sqrt{-g}d^{3}x^{i}$ and $V_{0}$ represents a sufficiently large time dependent three-volume.

The high temperature of the plasma led us to consider the
situation where only the squared magnetic field $H^2$ survives and the electric field is put to $E^2=0$. The
resulting averaged energy-momentum tensor is identified as a
perfect fluid with modified expressions for the energy density
$\rho$ and pressure $p$ given by

\begin{eqnarray}
\rho & = & \frac{H^{2}}{2}\left(1-8\, \alpha \, H^{2}\right),\label{ro EM nlinear}\\
p & = & \frac{H^{2}}{6}\left(1-40\, \alpha \,
H^{2}\right).\label{p EM nlinear}
\end{eqnarray}

This source violates the energy condition $\rho +
3 p\geqslant 0$, as in the models presented before. Nonetheless, it
is worth to call attention that, in this case, the pressure is isotropic.
We will look for a static solution in a spherical geometry given by
the metric \eref{metrica Escalar}. The field equations become:

\begin{eqnarray}
b' & = & \rho r^{2},\label{iso p b'}\\
\phi' & = & \frac{b+pr^{3}}{2r^{2}\left(1-\frac{b}{r}\right)},\label{iso p phi'}\\
p' & = & -\left(\rho+p\right)\phi'.\label{iso p p'}
\end{eqnarray}

Eliminating $ \phi'$ from these equations we obtain

\begin{equation}
\eqalign{b' = \rho r^{2},\cr
p' = -\left(\rho+p\right)\frac{\left(b+pr^{3}\right)}{2r^{2}\left(1-\frac{b}{r}\right)}.}
\label{sis iso}
\end{equation}

The boundary conditions at the wormhole's throat, $r_0$, are

\begin{eqnarray}
b_0\equiv  b(r_0)=r_0,\qquad & b'_0\equiv b'(r_0)\leqslant 1,\label{condcontb} \\
p_0\equiv p(r_0)=-\frac{1}{r_0^2},\qquad & p'_0\equiv p'(r_0)=\frac{b'_0-3}{r_0^3}<0, \label{condcontp}
\end{eqnarray}

\noindent which imply for $X(r)\equiv H^{2}(r)$:

\begin{equation}
X_0\equiv X(r_0)= \frac{1+\sqrt{1+\gamma}}{80\alpha}\geqslant
\frac{1}{16\alpha},\qquad X'_0\equiv X'(r_0)>0,
\label{condcontX}
\end{equation}

\noindent where $\gamma=960\alpha/r_0^2$.

The system \eref{sis iso} becomes

\begin{equation}
\eqalign{b' = \frac{X}{2}\left(1-8\alpha X\right) r^{2},\cr
X' = \frac{X\left(16\alpha X-1\right)\left(40\alpha X^{2}r-Xr-\frac{6b}{r^2}\right)}{\left(1-\frac{b}{r}\right)\left(80\alpha
X-1\right)},}
\label{sisX}
\end{equation}

\noindent and any solution of it with the above boundary conditions
has the property that the radial coordinate possesses a minimum
value at $r_0$. However, for this case, these boundary conditions
turn the solution completely unacceptable as we move away
from the throat. We can see this by examining the behavior of $X(r)$.
This function begins with a positive value and increases, so it must
reach a maximum value, $X_m\equiv X(r_m)>X_0$, at $r_m>r_0$ if we
want the functions $p(r)$, $\rho(r)$, $b(r)$ and $\phi(r)$ to remain
finite. Otherwise, the resulting space-time would not
be acceptable. Imposing that
$b(r)/r$ be finite (requirement of asymptotic flatness),
we can see by \eref{sisX} that $X(r)$ will only
stop increasing if one of these equations holds true:

\begin{equation}
\eqalign{X_{m} =0,\cr
16\alpha X_{m}-1 =0,\cr
40\alpha X_{m}^{2}r_{m}-X_{m}r_{m}-6\frac{b_{m}}{r_{m}^{2}}=0,}\label{raizesXl}
\end{equation}

\noindent with $b_m\equiv b(r_m)$.

It is clear that the first one cannot be true, for it would
imply that $X_m<X_0$. This same argument
discards the second one, which states that $X_m=1/16\alpha\leqslant
X_0$ and not greater. The last one gives

\begin{equation}
X_m=\frac{1+\sqrt{1+960\alpha\frac{b_m}{r_m^3}}}{80\alpha}=
\frac{1+\sqrt{1+\gamma\frac{r_0^2b_m}{r_m^3}}}{80\alpha} \label{Xm3},
\end{equation}

\noindent which is also smaller than $X_0$, since it is given by
\eref{condcontX}, and $b_m\leqslant r_m$ and $r_0<r_m$ implies that
$r_0^2b_m/r_m^3<1$.

Therefore, $X(r)$ diverge positively, leading the quantities
$p(r)$, $\rho(r)$, $b(r)$ and $\phi(r)$ to diverge negatively,
which would imply in a infinitely negative mass and a horizon for
large values of $r$. The only case in which $X(r)$ (and the other
functions) can stop increasing is if $b(r)$ increases in such a
way that $b(r)/r$ do also diverges and makes $X'(r)=0$, which is
not compatible with the requirement of asymptotic flatness. Hence
we can conclude that any attempt to impose a minimum value for the
radial coordinate in this case result in a too much problematic
space-time for it to be considered a satisfactory solution. This
case does not allow a reasonable wormhole solution.

\section{Conclusion}
In this paper we have investigated the existence of wormhole
solutions in theories which produce cosmological bounces.
We have shown that in one of them, concerning non-linear
electrodynamics, no wormhole solution is
possible, showing that it is not mandatory that violation of WEC, or theories
which produce cosmological bounces, can also result in wormholes.

In the other two theories we investigated, we were able to obtain
simple, traversable (in the sense of \cite{thorne}) wormhole
solutions with very nice properties. Both depend on a a parameter
$r_0$, which cannot be very big in order to not impose
unattainable speeds or a very long time to traverse the wormhole
(see equations \eref{convy} and \eref{convy2}).

In the case of section 2, non-minimal coupling between gravity and
electromagnetism, one arrives at the nice situation that no
electromagnetic energy (in the usual sense) is required to produce
the wormhole. The parameter $r_0$ appears only in the denominator
of the amplitude of the vector potential (see \eref{Sol Omega ANM} and the definition of $\Upsilon$). However, as there is
another arbitrary constant present in \eref{Sol Omega ANM},
one can adjust it to satisfy physical requirements while
preserving values of $r_0$ compatible with traversability with
reasonable speeds.

In the case of section 3, the parameter $r_0$ appears in the
energy density of the field (in the denominator), and in its total
mass (in the numerator). Hence, a small $r_0$ would imply a small
total quantity of scalar field, which is good, but a high
concentration of the field in the throat. However, if this field
does not interact with matter and other fields, the traveller
could pass through the throat without any harm.

Finally, we would like to point out that the solutions obtained here are all static,
as if those wormholes always existed. In order to see if they can really arise, it is
necessary to consider a dynamical geometry. The resulting equations and solutions should
then show if some initial configuration could evolve into a wormhole, and if it will be stable.
It should also be possible to investigate if during the bouncing period, where WEC is only preserved
when spacial sections have positive curvature, wormholes would have time to arise or not.

We have shown that the existence of static wormholes is indeed a perfectly plausible theoretical possibility,
in reasonable theoretical frameworks. Their production in the history of the universe or even in the laboratory
depend on dynamical developments of the present investigation, which will be the subject of our future investigations.

\ack We acknowledge CNPq and CAPES for financial support.


\section*{References}
\begin{thebibliography}{99}

\bibitem{acc} Perlmutter S \etal 1998 {\it Nature (London)} {\bf 391} 51
\nonum Riess A \etal 1998 {\it Astron. J.} {\bf 116} 1009

\bibitem{null} Barcel\'{o} C and Visser M 2002 {\it Int. J. of Mod. Phys.} D
{\bf 11} 1553

\bibitem{rip} Caldwell R R, Kamionkowski M and Weinberg N N 2003 {\it Phys. Rev. Lett.} {\bf 91} 071301

\bibitem{nonm} Fewster C J and Osterbrink L W 2006 {\it Phys. Rev.} D {\bf 74} 044021
\nonum Carvalho F C and Saa A 2004 {\it Phys. Rev.} D {\bf 70} 087302

\bibitem{qc} Nojiri S, Odintsov S D and Sami M 2006 {\it Phys. Rev.} D {\bf 74} 046004
\nonum Alam U and Sahni V 2006 {\it Phys. Rev.} D {\bf 73}, 084024
\nonum Pinto-Neto N and Santini E S 2003 {\it Phys. Lett.} A {\bf 315} 36

\bibitem{be} Pinto-Neto N and Fraga B M O Cosmic acceleration from interaction of ordinary fluids {\it Preprint} arXiv:0711.3602

\bibitem{inflation} Starobinsky A A 1979 {\it Pis'ma Zh. Eksp. Teor. Fiz.} {\bf 30} 719 [1979 {\it JETP Lett.} {\bf 30} 682]
\nonum Mukhanov V and Chibisov G 1981 {\it JETP Lett.} {\bf 33} 532
\nonum Guth A 1981 {\it Phys. Rev.} D {\bf 23} 347
\nonum Linde A 1982 {\it Phys. Lett.} B {\bf 108} 389

\bibitem{bounce} Tolman R C 1931 {\it Phys. Rev.} {\bf 38} 1758
\nonum Murphy G 1973 {\it Phys. Rev.} D {\bf 8} 4231
\nonum Melnikov V and Orlov S 1979 {\it Phys. Lett} A {\bf 70} 263
\nonum Barros J A, Pinto-Neto N and Sagioro-Leal M A 1998 {\it Phys. Lett.} A {\bf 241} 229
\nonum Colistete Jr. C, Fabris J C and Pinto-Neto N 2000 {\it Phys. Rev.} D {\bf 62} 083507

\bibitem{PP} Peter P and Pinto-Neto N 2002 {\it Phys. Rev.} D {\bf 65} 023513

\bibitem{thorne} Morris M S and Thorne K S 1988 {\it Am. J. of Phys.} {\bf 56} 395

\bibitem{Novello} Novello M and Salim J M 1979 {\it Phys. Rev.} D {\bf 20} 377

\bibitem{Novello1} Novello M, Oliveira L A R, Salim J M and Albaz E 1993 {\it Int. J. of Mod. Phys.} D {\bf 1} 641

\bibitem{PP2} Peter P and Pinto-Neto N 2002 {\it Phys. Rev.} D {\bf 66} 063509

\bibitem{Novello2} De Lorenci V A, Klippert R, Novello M and Salim J M 2002 {\it Phys. Rev.} D {\bf 65} 063501

\bibitem{Drummond} Drummond I T and Hathrell S J 1980 {\it Phys. Rev.} D {\bf 22} 343

\bibitem{prokopec} Prokopec T Cosmological magnetic fields from photon coupling
to fermions and bosons in inflation {\it Preprint} arXiv:astro-ph/0106247 and references therein
\nonum Prokopec T and Puchwein E 2004 {\it Phys. Rev.} D {\bf 70} 043004 and references therein

\bibitem{turner} Turner M S and Widrow L M 1988 {\it Phys. Rev.} D {\bf 37} 2743

\bibitem{slava} Golovnev A, Mukhanov V and Vanchurin V Vector Inflation {\it Preprint} arXiv:0802.2068

\bibitem{FPP} Finelli F, Peter P and Pinto-Neto N Spectra of primordial fluctuations in two-perfect-fluid regular bounces {\it Preprint} arXiv:0709.3074

\bibitem{BV} Bozza V and Veneziano G 2005 {\it Phys. Lett.} B {\bf 625} 177

\bibitem{Tolman} Tolman R C and Ehrenfest P 1930 {\it Phys. Rev.} {\bf 36} 1791

\end{thebibliography}
\end{document}